\documentclass[journal,twocolumn]{IEEEtran}
\usepackage{cite}
\usepackage[T1]{fontenc} 
\usepackage{amsmath,amssymb,amsfonts,amsthm}
\usepackage{algorithm}
\usepackage{algorithmic}
\usepackage{graphicx}
\graphicspath{{./visiofigures/}{./SimFigures/}{./BioFigures/}}
\usepackage{textcomp, epstopdf}
\usepackage{xcolor}
\usepackage{booktabs,multirow}
\usepackage{tabularx}
\usepackage[inline]{enumitem}
\usepackage{cite}
\usepackage{bigints, color}
\usepackage[unicode=true,
 bookmarks=false,bookmarksnumbered=true,bookmarksopen=false,bookmarksopenlevel=1,
 breaklinks=false,pdfborder={0 0 0},pdfborderstyle={},backref=false,colorlinks=false]
 {hyperref}
\hypersetup{pdftitle={Title},
 pdfauthor={Admin},
 pdfpagelayout=OneColumn, pdfnewwindow=true, pdfstartview=XYZ, plainpages=false}
 
\makeatletter

\theoremstyle{remark}
\newtheorem{rem}{Remark}
\makeatother
\allowdisplaybreaks
\begin{document}
\title{Integrated Sensing and Communications with Joint Beam Squint and Beam Split for Massive MIMO}
\author{Liangyuan~Xu and~Feifei~Gao 
\thanks{L.~Xu and F.~Gao  are with the Department of Automation, Tsinghua University, Beijing 100084, China, and also with Beijing National Research Center for Information Science and Technology (BNRist), Beijing 100084, China (email: \protect\href{mailto:xly18@mails.tsinghua.edu.cn}{xly18@mails.tsinghua.edu.cn}; \protect\href{mailto:feifeigao@ieee.org}{feifeigao@ieee.org}).}
}
\maketitle
\begin{abstract}
Integrated sensing and communications (ISAC) has attracted tremendous attention for the future 6G wireless communication systems.
To improve the transmission rates and sensing accuracy, massive multi-input multi-output (MIMO) technique is leveraged with large transmission bandwidth. 
However, the growing size of transmission bandwidth and antenna array results in the beam squint effect, which hampers the communications.
Moreover, the time overhead of the traditional sensing algorithm is prohibitive for practical systems.
In this paper, instead of alleviating the wideband beam squint effect, we take advantage of joint beam squint and beam split effect and propose a novel user directions sensing method integrated with massive MIMO orthogonal frequency division multiplexing (OFDM) systems.
Specifically, with the beam squint effect, the BS utilizes the true-time-delay (TTD) lines to steer the beams of different OFDM subcarriers towards different directions simultaneously.
The users feedback the subcarrier frequency with the maximum array gain to the BS.
Then, the BS calculates the direction based on the subcarrier frequency feedback.
Futhermore, the beam split effect introduced by enlarging the inter-antenna spacing is exploited to expand the sensing range.
The proposed sensing method operates over frequency-domain, and the intended sensing range is covered by all the subcarriers simultaneously, which reduces the time overhead  of  the conventional sensing significantly.
Simulation results have demonstrated the effectiveness as well as the superior performance of the proposed ISAC scheme.
\end{abstract}


\section{Introduction}
\IEEEPARstart{T}{he}  sixth generation (6G) wireless communication systems tend to enable  various emerging applications such as virtual reality, artificial intelligence (AI), intelligent manufacturing,  autonomous driving and so on \cite{6G1,6G2,6G3}.
On the one hand, these applications demand ultra-reliable low-latency links and higher transmission rates for communications.
On the other hand, these applications impose stringent requirements on high-resolution and robust sensing  ability.
Hence, sensing has also been recognized as the indispensable functionality for future mobile communication systems, which motivates the studies of sensing
and communications (S\&C) \cite{SC1,SC2,SC3}.

Sensing can be operated in a passive or an active mode \cite{KHarray,CJCsource}.
In passive sensing, the information (e.g., location, AoD, velocity)  is estimated from the wireless signal transmitted/reflected by the target device \cite{ABpassive,LJLpassive,Atomic,XLYonebit}.	
In the active sensing mode, the active transmitter (radar or active sonar) emits a known waveform and then estimates some unknown parameters (e.g., AoD, speed, bearing) from the reflected echo \cite{phasedradar,MIMOradar,LJmimoradar}.
Owing to the special demands for hardware platform, the main application of the active sensing technique is for radar.
Since the  waveform, the hardware architecture and the frequency band  of communication systems are different from radar, the active sensing  technique can not be directly adopted in communication systems.
Sensing and communications are individually accomplished by different systems without effective collaboration and resource sharing.
Therefore, the fusion of wireless communications and radar sensing recently gains growing attention.
In \cite{radcom}, the authors have designed a radar-communications (RadCom) system. 
The authors of \cite{JCR} have developed a joint communication-radar (JCR) proof-of-concept platform.
The authors in \cite{DFRC} have devised a dual-functional radar-communication (DFRC) system.
Moreover, the demand for the fusion of sensing and communications motivates more recent researches about
integrated sensing and communications (ISAC) \cite{ISAC1,ISAC2,ISAC3,ISAC4,ISAC5}.
Nevertheless, the major obstacle of ISAC is how to implement sensing and communications with shared spectrum/hardware resources and effective time/spatial/frequency multiplexing technology.

By deploying a massive antenna array with hundreds or thousands of elements at the base station (BS), the spectral efficiency and sensing accuracy can be improved significantly \cite{massive}.
Furthermore,  to fulfill increasing demands for higher transmission rates and accommodate more devices,  large transmission bandwidth will be adopted to collaborate with the massive MIMO technique \cite{mmWave}.
However, one major issue for ISAC is the beam squint effect in wideband massive MIMO system. 
Specifically, the wideband transmission and the large antenna array yield  non-negligible signal propagation delay across the
antenna array  in one sample period, which is called \emph{wideband effect} or \emph{beam squint effect} \cite{chirpfiber,CMMsquint,WBLmag}.
The beam squint effect makes  the analog beamforming direction vary as a function of the frequency, which brings some obstacles for traditional narrowband algorithms \cite{WBL}. 
In the literature, there have been some works referring to the beam squint effect, e.g.,  channel estimation \cite{WBLce,WMJsquint,JMNsquint,tensorYXH,MSQsquint,Mathewsquint}, performance analysis \cite{analySquint,NSsquint}, and precoding/beamforming \cite{WBLthz,Hanzosquint,FJsquint}.
In \cite{WBLce}, taking beam squint effect into consideration, the authors have proposed a compressed sensing based  channel estimation algorithm. This algorithm performs traditional narrowband method for the frequency-division duplex (FDD)
 massive MIMO systems.
The authors of \cite{tensorYXH} have addressed the channel estimation problem for the MIMO orthogonal frequency division multiplexing (OFDM) systems with wideband effect.
The  work \cite{analySquint} has shown that the traditional codebook design neglects the beam squint effect and suffers from severe performance degradation when the bandwidth becomes sufficiently large.
In \cite{Mathewsquint} and \cite{WBLthz}, the authors have utilized the true-time-delay (TTD) lines to mitigate the beam squint effect in massive MIMO terahertz (THz) systems. 
The authors in  \cite{FJsquint} have devised a new precoder/combiner to form a radiation pattern with a wide beam, which can alleviate the beam squint effect in sub-THz MIMO-OFDM systems.
Though all these works tend to eliminate  the beam squint effect, the beam squint effect can actually be exploited to facilitate sensing and communications.
Since the beamforming direction will change as frequency varies owing to the beam squint effect, the beam sweeping can be performed by varying frequency, and then users in different directions can be sensed.
However, very few results are available about the beam squint issue in ISAC, especially in sensing. 

In fact, another issue for ISAC is the complexity and the time overhead.
Traditional sensing method is the beam training, e.g., the exhaustive searching to find the user directions. 
However,  the overall search time is prohibitive for practical application. 
Hence, some works have been investigated to improve the search efficiency for the beam training.
For example, a hierarchical search method has been adopted in \cite{WJtraining,DJtraning,HTtraining,LCtraining,WJHtraining}. 
First, a coarse codebook with low-resolution is used to find the coarse direction range, and then a fine search is applied to the best high-resolution direction.
This hierarchical search method can reduce the searching overhead significantly.
Moreover, the authors of \cite{XZYtraining} have exploited sub-array technique and devised a binary-tree search method to improve the beam training efficiency.
Furthermore, the authors in \cite{NBYtraining} utilize partial search as well as hierarchical search and developed a ternary-tree based beam search method, which further reduces  beam search time.
Nevertheless, these methods therein have been operated over different time slots, which sequentially and individually tests all possible beam directions to find the true direction.
Moreover, these methods are intrinsically time-consuming, and it is hard to reduce the time overhead by orders of magnitude.

In this paper, under the assumption of wideband transmission, we propose two types of user directions sensing methods with joint beam squint and beam split effect for massive MIMO-OFDM systems.
The contributions of this work are summarized as follows:
\begin{itemize}
\item 
The beam squint effect from wideband massive MIMO has been considered for the sensing methods. 
Unlike the existing works that intend to mitigate the beam squint effect, we take advantage of the wideband beam squint effect instead. 
By deploying TTD lines, we make the beam squint range adjustable and sufficiently wide with a fixed bandwidth, such that the BS can  cover the whole range of all possible directions. 

\item 
By increasing the inter-antenna spacing (e.g., larger than half the wavelength), we introduce the \emph{beam split effect} and exploit it during user directions sensing.
With joint beam squint and beam split, the sensing range will split into several ranges, which expands the sensing range compared to the method with beam squint only.
Since the users in different splitted ranges will be sensed by the same subcarrier, directions sensing ambiguities occur with beam split.
We then propose a  \emph{intersection validation} method to alleviate the ambiguities and determine the true direction.   

\item 
Different from the traditional beam sweeping/training method that performs over different time slots, the proposed sensing method operates over frequency-domain.
With beam squint effect, the beams of different OFDM subcarriers steer towards different directions.
Hence, the users in the sensing range will be covered by the beams of the subcarriers within only one OFDM block simultaneously.
The proposed user directions sensing method will reduce the time overhead significantly.

\item
The proposed user directions sensing method is integrated with communication systems. 
The BS emits beams of different subcarriers towards different directions simultaneously.
The users feedback the subcarrier frequency with the maximum array gain to the BS.
The BS then calculates the user direction according to the subcarrier frequency feedback.
The proposed sensing method shares spectrum/hardware resources with general communication systems, Only with some feedbacks from users to the BS.
The proposed sensing method shows superiority for the application of ISAC.
 
\end{itemize}

The remainder of this paper is organized as follows.  
The system model, the wideband beam squint effect as well as the beam split effect are introduced in Section \ref{sec:model}. 
Section \ref{sec:sensingmethod} illustrates the proposed user directions sensing method with beam squint only as well as  the another proposed sensing algorithm with joint beam squint and beam split. 
Numerical results are provided in Section \ref{sec:sim}.
Conclusions are finally made in Section \ref{sec:con}.

\noindent\textbf{Notation:} 
Uppercase boldface $\mathbf{X}$, lowercase boldface $\mathbf{x}$ and lowercase non-bold $x$ denote matrices, vectors and scalars, respectively. 
Superscripts $(\cdot)^T$,  $(\cdot)^H$ and $(\cdot)^{*}$ represent the transpose, the Hermitian transpose and the complex conjugate, respectively. 
We use $\|\cdot\|_p$ for $\ell_p$-norm.  
Symbols $\cap$ denote the intersection of two set. 
The real and imaginary parts of complex numbers are denoted by $\Re(\cdot)$ and $\Im(\cdot)$, respectively. 
$\jmath = \sqrt{-1}$.

\section{Communication Model of Massive MIMO} \label{sec:model}
\begin{figure}[!tpb]
\centering
\includegraphics[width=0.46\textwidth]{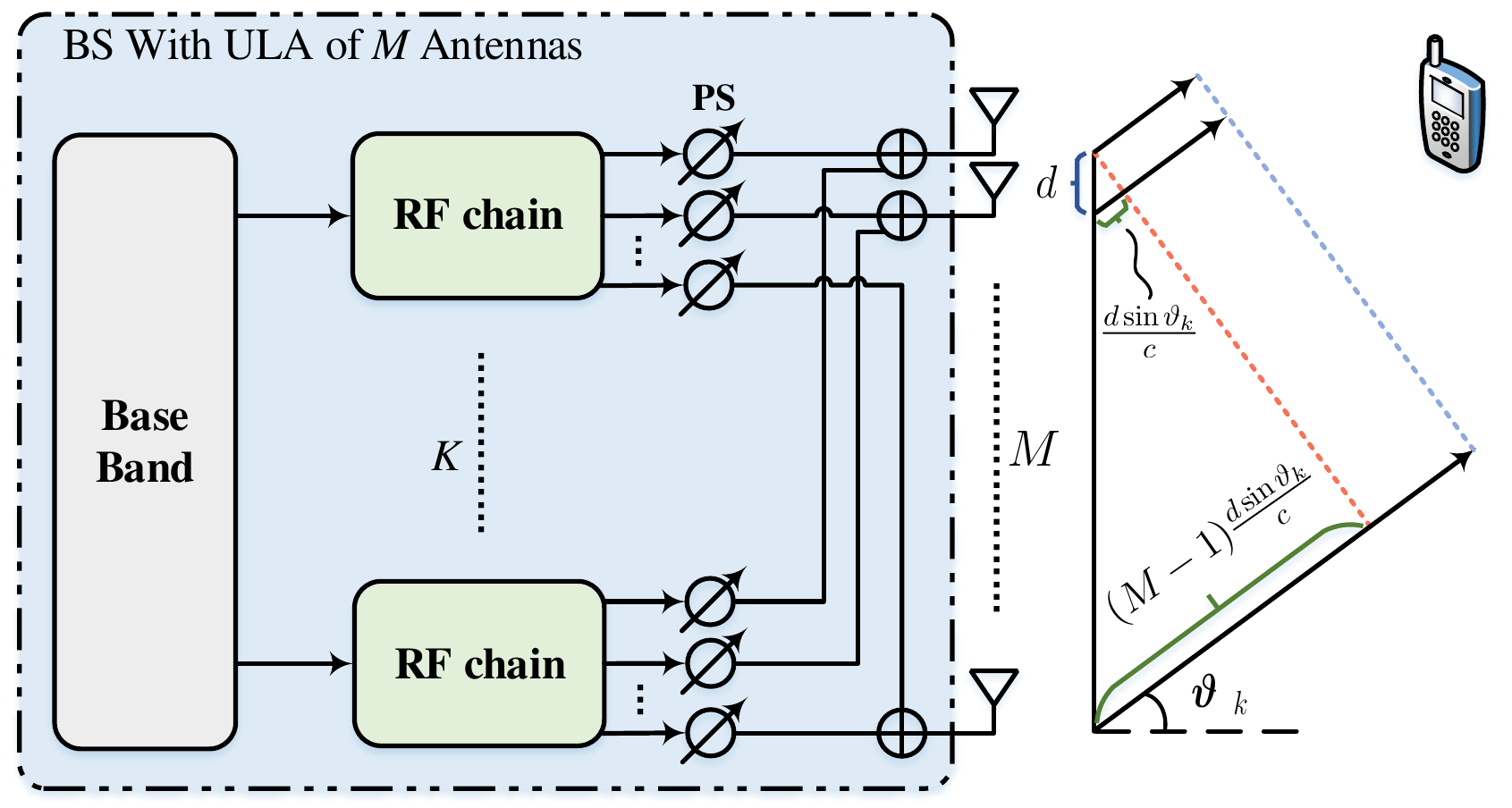}
\caption{Massive MIMO system with a ULA antenna array and the signal propagation delay across the array aperture.}
\label{fig:systdiagram}
\end{figure}

Consider a multi-user mmWave massive MIMO system where $K$ single antenna users are served by a BS. 
As illustrated in Fig. \ref{fig:systdiagram}, a hybrid architecture with total $K$ RF chains is deployed at the BS, and each RF chain serves one user.
Additionally, the BS is equipped with a uniform linear array (ULA)\footnote{In this paper, we only consider ULA to derive the beam squint and beam split effects. Nevertheless, the results of ULA can be extended to the case of uniform planar array (UPA) by dividing UPA into ULAs with two dimensions.} of $M$ antennas with inter-antenna spacing $d$. 
The transmission bandwidth and carrier frequency are denoted by $F$ and $f_c$, respectively.
We consider a  line-of-sight (LOS) propagation  scenario that each user has only a LOS path to the BS\footnote{For mmWave and sub-THz communications, LOS propagation plays a major role owing to the extreme penetration loss, high attenuation and low diffractions \cite{RappLOS}. Nevertheless, the proposed method can be extended to the multi-path case.}. 
Under the far-field assumption that the BS array aperture is much smaller than the distance between users and the BS, a planar wave is transmitted between the BS and users.
The angle of departure (AoD) of the $k$th user's LOS path is denoted as $\vartheta_{k}$ that indicates the user location information\footnote{With multi-BS cooperation, the location of the user can be obtain from the angles.}.
Then, the propagation delay $\tau_{k, m}$ from the $m$th antenna of the BS to user-$k$ can be written as \cite{WBL}
\begin{equation}\label{eq:delay}
\tau_{k, m} = \tau_{k,1}+(m-1) \cdot \frac{d \sin \vartheta_{k}}{c}=\tau_{k,1}+(m-1) \cdot \frac{\psi_{k}}{f_{c}},
\end{equation}
where $\tau_{k,1}$ is the propagation delay from user-$k$ to the first antenna of the BS, $c$ is the speed of the light, and $\psi_{k} \triangleq \frac{d \sin \vartheta _{k}}{\lambda _c} $ is defined as the normalized AoD. The  passband signal received by  user-$k$ from the $m$th antenna of the BS can be represented as 
\begin{equation}\label{eq:passband}
\begin{aligned}
&\Re\{\alpha_{k} s_{k,m}(t-\tau_{k, m}) e^{j2\pi f_{c} (t-\tau_{k, m})} \} \\
& = \Re\{\alpha_{k} s_{k,m}(t-\tau_{k, m}) e^{-j2\pi f_{c} \tau_{k,1}} e^{-j2\pi (m-1)\psi_{k}} e^{j2\pi f_{c} t} \},
\end{aligned}
\end{equation}
where $\alpha_{k}$ is the complex channel gain, and $s_{k,m}(t)$ is the  baseband signal transmitted by the $m$th antenna of the BS. Hence, the corresponding baseband signal received by  user-$k$ from the $m$th antenna of the BS is 
\begin{equation}\label{eq:baseband}
\begin{aligned}
& \alpha_{k} s_{k,m}(t-\tau_{k, m}) e^{-j2\pi f_{c} \tau_{k,1}} e^{-j2\pi (m-1)\psi_{k}} \\
& = \beta_{k} s_{k,m}(t-\tau_{k, m})  e^{-j2\pi (m-1)\psi_{k}},
\end{aligned}
\end{equation}
where $\beta_{k} \triangleq \alpha_{k} e^{-j2\pi f_{c} \tau_{k,1}}$ is defined as the \emph{equivalent baseband complex gain}. Then, the \emph{time-domain} downlink channel between the $m$th antenna of the BS and user-$k$ can be expressed as 
\begin{equation}\label{eq:timedomain}
\begin{aligned}
h_{k,m}(t) & = \beta_{k} e^{-j2\pi (m-1)\psi_{k}} \delta(t-\tau_{k, m})  \\
& = \beta_{k} e^{-j2\pi (m-1)\psi_{k}} \delta(t-\tau_{k,1}-(m-1) \frac{\psi_{k}}{f_{c}}),
\end{aligned}
\end{equation}
where $ \delta(\cdot) $ denotes the Dirac delta function.

By taking the Fourier transform of $h_{k,m}(t)$, we obtain the \emph{frequency-domain} channel response as 
\begin{equation}\label{eq:frequencydomain}
\begin{aligned}
h_{k,m}(f) & = \beta_{k} e^{-j2\pi (m-1)\psi_{k}} e^{-j2\pi f \tau_{k, m}}  \\
& = \beta_{k} e^{-j2\pi (m-1)\psi_{k}(1+\frac{f}{f_c})} e^{-j2\pi f \tau_{k, 1}},
\end{aligned}
\end{equation}
where $f\in[0,F]$. Let us define $\Theta(f) \triangleq \psi_{k}(1+\frac{f}{f_c})$.
Then, the \emph{spatial steering vector} corresponding to  the angle $\vartheta_{k}$ can be written as 
\begin{equation}\label{eq:steering}
\mathbf{a}(\Theta(f))  = [1, e^{-j2\pi \Theta(f)}, \cdots, e^{-j2\pi (M-1)\Theta(f)} ]^{T}.
\end{equation}
By collecting \eqref{eq:frequencydomain} into a vector, we obtain the frequency-domain channel  between the BS and user-$k$ as
\begin{equation}\label{eq:channel}
\mathbf{h}_{k}(f)  = \beta_{k} e^{-j2\pi f \tau_{k, 1}} \mathbf{a}(\Theta(f)).
\end{equation}
Hence, the signal received by user-$k$ at frequency $f$ is given by
\begin{equation}\label{eq:received}
\begin{aligned}
y_{k}(f) & = \sum\limits^{M}_{m=1} h_{k,m}(f) s_{k,m}(f) + \nu \\
& = [h_{k,1}(f),\cdots,h_{k,M}(f)] [s_{k,1}(f),\cdots,s_{k,M}(f)]^{T} + \nu \\
& = \mathbf{h}_{k}(f)^{T} \mathbf{s}_{k}(f) + \nu,
\end{aligned}
\end{equation}
where $s_{k,m}(f)$ is the transmitted frequency-domain signal by the $m$th antenna of the BS, and $\nu$ is the additive Gaussian noise with zero mean and variance $\sigma^2$.

For the massive MIMO-OFDM system with $N$ subcarriers and the subcarrier spacing as $\frac{F}{N}$, the signal received by the $k$th user at the $n$th carrier is 
\begin{equation}\label{eq:OFDM}
y_{k}(nF/N) = \mathbf{h}_{k}(nF/N)^{T} \mathbf{s}_{k}(nF/N) + \nu,
\end{equation}
where $\mathbf{s}_{k}(nW/N)$ is the transmitted signal vector at the $n$th carrier.

\subsection{The Large Array Wideband Beam Squint Effect}\label{sec:squint}
According to \eqref{eq:delay}, the maximum time delay difference $\Delta \tau_{k}$ among the antenna array is $\Delta \tau_{k}=(M-1) \frac{d \sin \vartheta_{k}}{c}$. 
Under traditional narrow bandwidth assumption, the symbol period $T_{s}$ of the transmitted signal $s_{k,m}(t)$ is much longer than $\Delta \tau_{k}$ ($T_{s}\gg \Delta \tau_{k}$), which indicates that $s_{k,m}(t-\tau_{k, m})\approx s_{k,m}(t-\tau_{k,1})$. Then, the spatial steering vector in \eqref{eq:steering} reduces to 
\begin{equation}\label{eq:steeringnarrow}
\mathbf{a}(\psi_{k})  = [1, e^{-j2\pi \psi_{k}}, \cdots, e^{-j2\pi (M-1) \psi_{k}} ]^{T},
\end{equation}
which is widely adopted under mmWave narrow bandwidth assumption \cite{xhxTVT}.  To maximize the antenna array gain, the optimal beamforming vector under this narrowband assumption is chosen as $\mathbf{w}_{k}=\mathbf{a}(\psi_{k})^{*}$, which yields the maximum array gain as 
\begin{equation}
\left | \mathbf{a}(\psi_{k})^{T} \mathbf{w}_{k} \right | = M.
\end{equation}
For the hybrid analog/digital architecture, the value of the PSs are  set according to $\mathbf{w}_{k}$ to employ the analog beamforming with maximum gain \cite{WBL}.

Note that, the maximum time delay difference $\Delta \tau_{k}$ is proportional to the array size $M$. Hence, under the assumption of wideband and large array, the approximation $s_{k,m}(t-\tau_{k, m})\approx s_{k,m}(t-\tau_{k,1})$ does not hold. Then, the spatial steering vector $\mathbf{a}(\Theta(f))$ is the same as \eqref{eq:steering}.
We note here that $\mathbf{a}(\Theta(f))$ is frequency-dependent, which indicates that the steering vector will change as frequency varies, although the AoD $\vartheta_{k}$ is fixed. 
Therefore, to obtain the maximum array gain $M$, the analog beamforming direction  should also change as frequency varies. 
This is named as \emph{wideband beam squint effect}. 
However, for the PS-based analog beamforming, the phase of PSs is usually unchanged in the whole bandwidth since the PS is a narrowband frequency-independent device. 
Assuming we intend to implement beamforming towards the angle $\phi_{k}$, the phase of the $M$ PSs should be chosen as $\mathbf a ( \phi_{k}  )^{*}$, which yields
\begin{equation}\label{eq:PSnarrowband}
\mathbf{w}_k= \mathbf a ( \phi_{k}  )^{*} = [1,e^{j2\pi \phi_{k}    },\dots,e^{j2\pi (M-1) \phi_{k}    } ]^T.
\end{equation}
Then, under wideband assumption, the array gain at frequency $f$ is 
\begin{equation}\label{eq:gainwideband}
\left|\mathbf{a}^{T}\left(\Theta(f)\right) \mathbf{w}_{k}\right| =\left|\sum_{m=1}^{M} e^{-j 2 \pi(m-1) (1+\frac{f}{f_c}) \psi _{k}} \cdot e^{j 2 \pi(m-1) \phi_{k}   } \right|. 
\end{equation}
The array gain in \eqref{eq:gainwideband} achieves the maximum value $M$ when the following condition is satisfied
\begin{equation}\label{eq:squintrule}
\psi _{k}  = \frac{\phi_{k} }{1+\frac{f}{f_c} }.
\end{equation}
From \eqref{eq:squintrule}, we note that even though the intended beamforming angle is $\phi_{k}$, the matched normalized AoD $\psi _{k}$ with maximum array gain will decrease as the frequency $f$ increases.
Let us define the \emph{beam squint range} as the maximum range of the normalized AoD variation within the transmission bandwidth $F$, which can be written as 
\begin{equation}\label{eq:squintrange}
\Delta \psi _{k}  = \phi_{k}\left(1 - \frac{ 1 }{1+\frac{F}{f_c} } \right) = \phi_{k} \frac{ F }{f_c + F}.
\end{equation}
This wideband beam squint effect means that we can use a set of fixed PSs $\mathbf{w}_k$ to implement beamforming with frequency-dependent varying directions. 
For massive MIMO-OFDM communication systems, this beam squint effect causes beamforming performance degradation since subcarriers can not focus towards the same direction.
Thus, the compensation algorithms have been developed in \cite{WBLthz,Hanzosquint,FJsquint} to mitigate the beam squint effect. 
Nevertheless, if the beam squint effect is exploited  rather than eliminated in the sensing scheme, then we can use a same analog beamforming vector to steer different subcarriers towards different directions simultaneously. 
Then, users in different directions can be covered and sensed simultaneously, which reduces the time overhead of sensing algorithms significantly.

\begin{figure}[!tpb]
\centering
\includegraphics[width=0.46\textwidth]{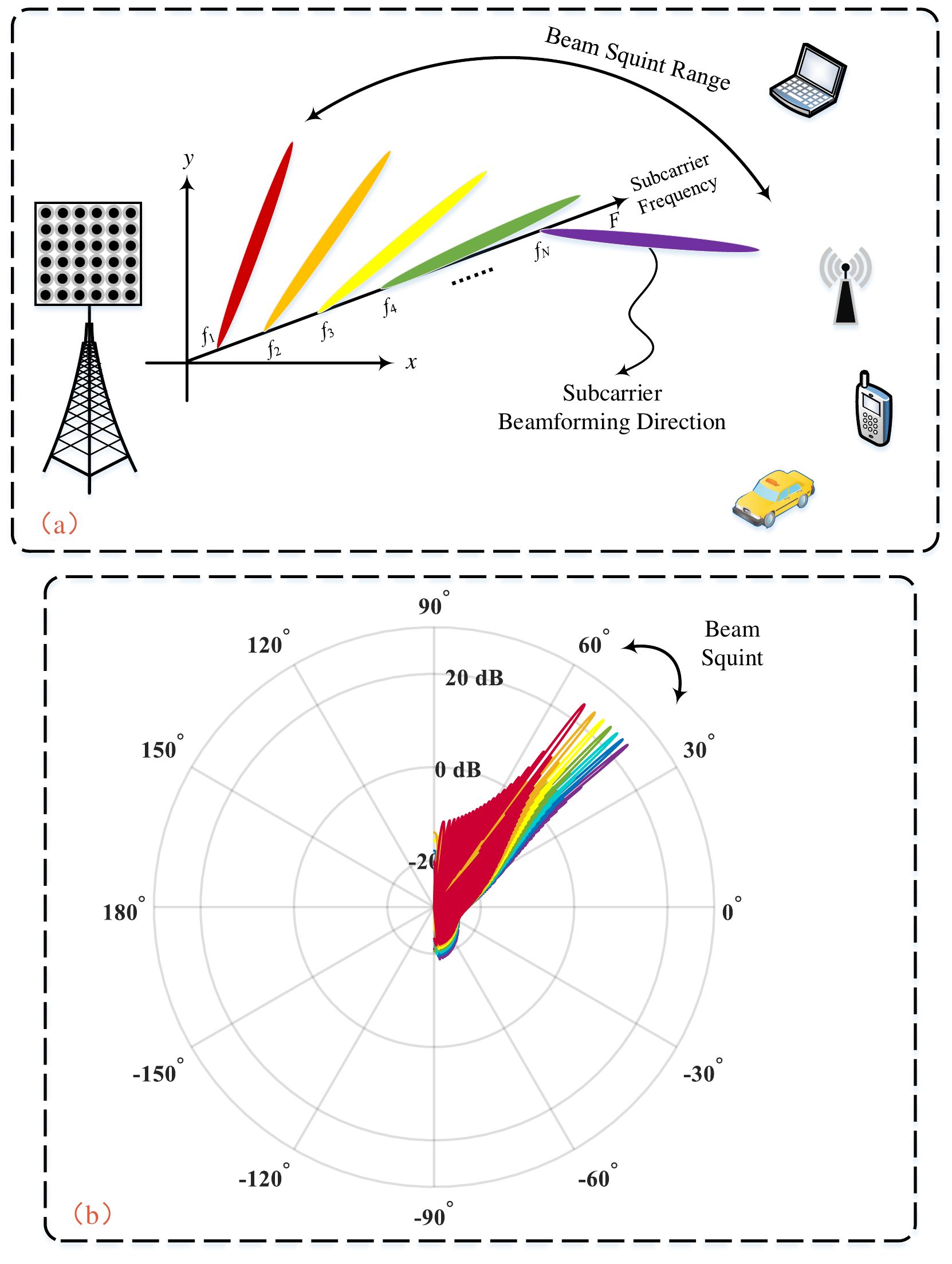}
\caption{(a) The demonstration of the wideband beam squint effect of massive MIMO system. (b) The beam pattern of one RF chain with fixed PSs, where different color denotes the beam of different subcarrier.}
\label{fig:beamsquint}
\end{figure}

Fig. \ref{fig:beamsquint} illustrates the wideband beam squint effect of massive MIMO system. Fig. \ref{fig:beamsquint}(a) shows the beamforming direction of one RF chain with a set of fixed PSs.  As a consequence of beam squint effect, the beam direction of different subcarrier varies as subcarrier frequency increases. 
Fig. \ref{fig:beamsquint}(b) demonstrates the beam pattern of one RF chain with fixed PSs. 
We adopt different color in  Fig. \ref{fig:beamsquint}(b) to distinguish the beam pattern at different subcarrier frequency, which is so called `rainbow' pattern under beam squint effect. 
It is worth noting that we set the bandwidth as $F=0.2f_{c}$ in  Fig. \ref{fig:beamsquint}(b).
Although the bandwidth is sufficiently wide,  the beam squint range is nevertheless not adequate for sensing purpose. 
Hence, for better user directions sensing, the  beam squint range should be adjustable with a limited bandwidth.

\subsection{The Beam Split Effect}\label{sec:split}
In traditional array signal processing, the inter-antenna spacing $d$ is required to be no larger than half the wavelength $\frac{\lambda_{c}}{2}$ in order to avoid phase ambiguity. However, if the  inter-antenna spacing is larger than $\frac{\lambda_{c}}{2}$ (especially a multiple of $\frac{\lambda_{c}}{2}$), then the phase ambiguity  occurs as a consequence of the Vandermonde structure of  the spatial steering vector. 
Specifically, we consider that the inter-antenna spacing is $d=P\frac{\lambda_{c}}{2}$ where $P$ is a positive number. 

For the narrowband case, the spatial steering vector corresponding to  the angle $\vartheta_{k}$ is 
\begin{equation}
\begin{aligned}
\mathbf{a}(\psi_{k})  &= [1, e^{-j2\pi \psi_{k}}, \cdots, e^{-j2\pi (M-1) \psi_{k}} ]^{T} \\
&= [1, e^{-j \pi P \sin \vartheta_{k}}, \cdots, e^{-j \pi (M-1) P \sin \vartheta_{k}} ]^{T}.
\end{aligned}
\end{equation}
Assume that  the steering vector $\mathbf{a}(\psi_{k}^{s})$ corresponding to another different angle $\vartheta_{k}^{s}$ has the same phase with $\mathbf{a}(\psi_{k})$. 
Since the $m$th element of the steering vector equals to the $(m-1)$th power of the second element, $\mathbf{a}(\psi_{k}^{s})$ has the same phase with $\mathbf{a}(\psi_{k})$ if and only if  the second element of both $\mathbf{a}(\psi_{k}^{s})$ and $\mathbf{a}(\psi_{k})$ should have the same phase, which yields
\begin{equation}\label{eq:2ndsteering}
e^{-j \pi P \sin \vartheta_{k}}  = e^{-j \pi P \sin \vartheta_{k}^{s}} = e^{-j \pi P \sin \vartheta_{k}^{s} + 2Z\pi},
\end{equation}
where $Z$ is an integer. For given $\vartheta_{k}$, we can obtain the corresponding $\vartheta_{k}^{s}$ from 
\begin{equation}\label{eq:split}
\begin{aligned}
& \pi P \sin \vartheta_{k}  = \pi P \sin \vartheta_{k}^{s}  + 2 Z \pi, \\
& Z = \pm 1, \pm 2, \pm 3, \cdots 
\end{aligned}
\end{equation}
In \eqref{eq:split},  we note that $\pi P(\sin \vartheta_{k} -\sin \vartheta_{k}^{s}) \in [-2\pi P , 2\pi P]$ and $2 Z \pi \in \{\pm 2\pi, \pm 4\pi, \pm 6\pi, \cdots\}$. Hence, the angle $\vartheta_{k}^{s}$ has different value with $\vartheta_{k}$ only if $P>1$. This can be named as \emph{beam split effect} for the narrowband case,  i.e., the steering vector $\mathbf{a}(\psi_{k})$ corresponding to  the angle $\vartheta_{k}$ also match other different angle $\vartheta_{k}^{s}$. Hence,  the analog beam will split to multiple different directions.  

For wideband case, the spatial steering vector corresponding to  the angle $\vartheta_{k}$ is given in \eqref{eq:steering} as $\mathbf{a}(\Theta(f))$. 
Similar to the narrowband case, we assume that  the steering vector $\mathbf{a}(\Theta(f)^{s})$ corresponding to another different angle $\vartheta_{k}^{s}$ has the same phase with $\mathbf{a}(\Theta(f))$, and we can obtain $\vartheta_{k}^{s}$ from the second element of both $\mathbf{a}(\Theta(f)^{s})$ and $\mathbf{a}(\Theta(f))$, which yields
\begin{equation}\label{eq:splitwideband}
\begin{aligned}
& \pi P \sin \vartheta_{k}(1+\frac{f}{f_c})  = \pi P \sin \vartheta_{k}^{s}(1+\frac{f}{f_c})  + 2 Z \pi, \\
& Z = \pm 1, \pm 2, \pm 3, \cdots 
\end{aligned}
\end{equation}
The \emph{beam split effect} for the wideband case is similar with the narrowband case that the analog beam will split to multiple different directions at any subcarrier. 
While the main difference is that the splitted beam directions $\vartheta_{k}^{s}$ will change as the frequency varies. 

\begin{rem}
Even though both the beam squint and the beam split effect can result in multiple different directions, the characteristics are  distinct.
For the beam squint effect, beam direction changes as the frequency varies, while at each subcarrier there is only one beam direction.
Hence the beams gradually `squint' aside as the frequency changes. 
For the beam split effect, multiple different directions occur at the same subcarrier, which is then named with `split'.
\end{rem}

\begin{figure}[!tpb]
\centering
\includegraphics[width=0.46\textwidth]{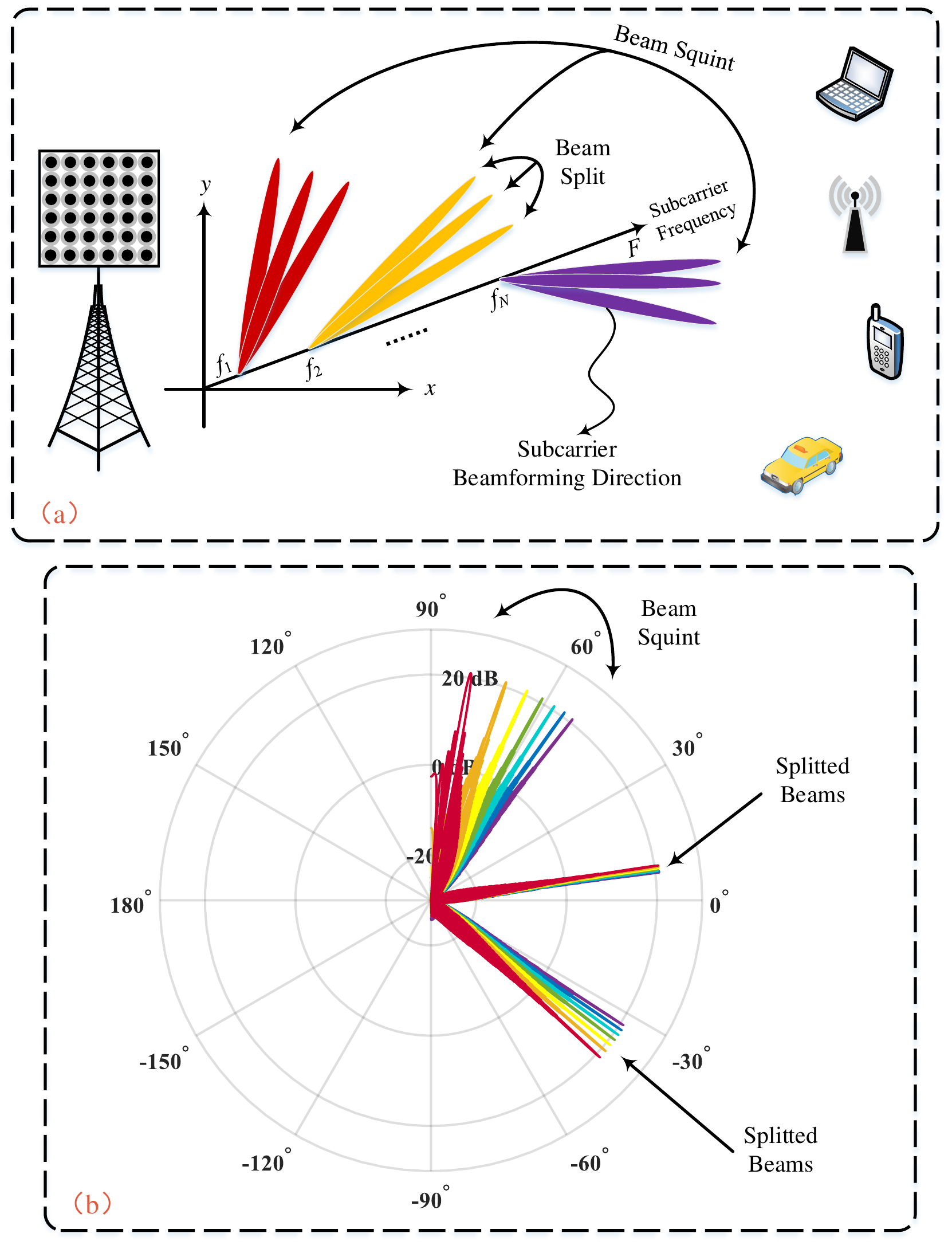}
\caption{(a) The illustration of the joint beam squint and split effect of massive MIMO system. (b) The beam pattern of one RF chain with fixed PSs, where different color denotes the beam of different subcarrier.}
\label{fig:beamsquintsplit}
\end{figure}

Fig. \ref{fig:beamsquintsplit}(a) illustrates the joint wideband beam squint and beam split effect.  
We can find that each subcarrier has several splitted beamforming directions and these angles will vary as subcarrier frequency changes.  
Fig. \ref{fig:beamsquintsplit}(b) shows the beam pattern of one RF chain with a set of fixed PSs, and different color denotes the beam pattern of different subcarrier frequency.
Note that in Fig. \ref{fig:beamsquintsplit}(b), we set the inter-antenna spacing as $d=\frac{3\lambda_{c}}{2}$, and the transmission bandwidth is chosen as $F=0.2f_{c}$.
In this case, there are no overlaps among the splitted beam squint ranges.
Nevertheless, if the beam squint range becomes wider and the number of splitted directions increases, then the splitted ranges will overlap and cause sensing  ambiguities.
Hence, the issue of overlap should be considered to avoid the ambiguity.
Additionally, it can be found in Fig.~\ref{fig:beamsquintsplit}(b) that the beam squint range of each splitted beam is not uniform.
One reason is that the bandwidth will influence the beam split effect in \eqref{eq:splitwideband},  and the other reason is that the sine function is a nonlinear function.  
Similarly, we should make the splitted beam squint range adjustable for better user directions sensing.

\section{Sensing With Beam Squint and Beam Split}\label{sec:sensingmethod}
For massive MIMO-OFDM system, by adopting wideband beam squint and beam split effect, we can make different subcarriers focus towards different directions simultaneously to cover a specific range.  
The users in the range feedback the subcarrier index with the maximum array gain to the BS. 
Then, the BS calculates the direction of the corresponding user based on the subcarrier index feedback. 
The beam sweeping and user directions sensing are achieved by one OFDM block in frequency-domain.
Compared with traditional beam sweeping that performs over different time slots, this sensing method with beam squint and beam split operates over frequency-domain and could reduce time overhead significantly since a wide range is covered by one OFDM block.
Therefore, the range of the subcarriers beamforming directions should be flexible and wide to cover the directions of all possible users. 
However, from \eqref{eq:squintrange}, it can be found that the beam squint range depends on the transmission bandwidth $F$. 
Hence, we should devise methods to adjust the beam squint range with a limited transmission bandwidth.

\begin{figure}[!tpb]
\centering
\includegraphics[width=0.46\textwidth]{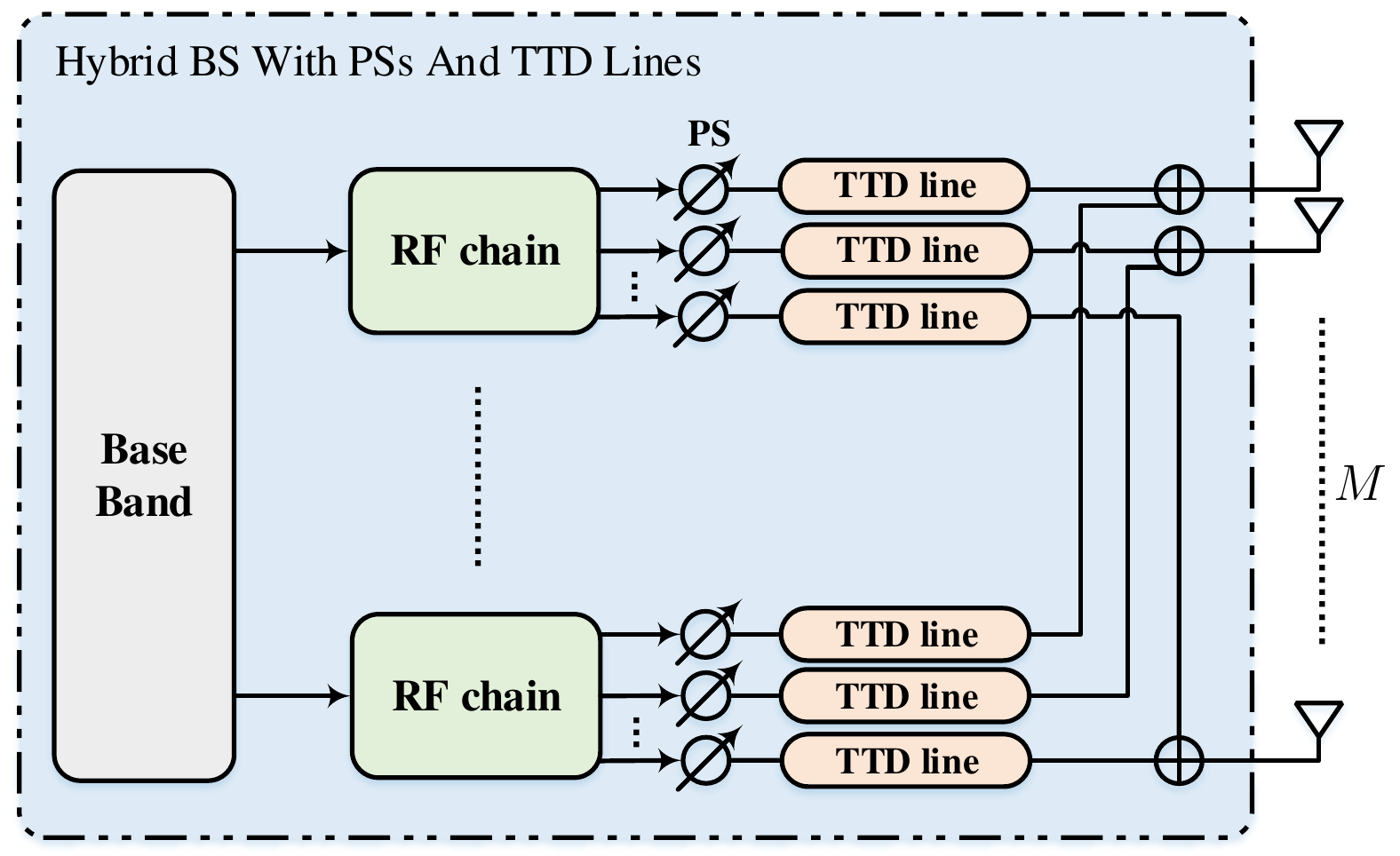}
\caption{Hybrid massive MIMO system with TTD lines are deployed between the PSs and the antennas.}
\label{fig:systdiagramTTD}
\end{figure}

As discussed in Section \ref{sec:squint}, the beam squint effect essentially results from the signal propagation delay across the antenna array. Hence, an effective method to make the beam squint range adaptable is to modify the propagation delay. Specifically, as shown in Fig. \ref{fig:systdiagramTTD}, we introduce TTD lines into the analog part of the transceivers to modify the signal propagation delay\footnote{In the literature, TTD lines are utilized to alleviate the delay for communications. Nevertheless, TTD lines are adopted here to increase/reduce the delay for ISAC.}. The TTD lines are deployed between the PSs and the antennas, and each PS is connected to a TTD line. Since the TTD line is a wideband device with frequency-dependent response, the combination of the PS and the corresponding TTD line is also a wideband device. 
Assume that  the $M$ PSs connected to the $k$th RF chain are chosen as $\mathbf a ( \phi_{k}  )^{*}$ to implement beamforming towards the angle $\phi_{k}$. 
Then, the combination of the $m$th TTD line and the $m$th PS connected to the $k$th RF chain has time-domain response as 
\begin{equation}\label{eq:PSTTDtime}
e^{j2\pi (m-1) \phi_{k}}\cdot\delta(t-t_{k,m}),
\end{equation}
where $t_{k,m}$ is the time delay introduced by the $m$th TTD line. Then, the frequency-domain response of \eqref{eq:PSTTDtime} is 
\begin{equation}\label{eq:PSTTDfrequency}
g_{k,m} = e^{j2\pi (m-1) \phi_{k}}\cdot e^{-j2\pi f t_{k,m}}.
\end{equation}
We can obtain the frequency response vector of the $k$th RF chain as $\mathbf{g}_{k} = [g_{k,1},g_{k,2},\cdots,g_{k,M}]^{T}$.

\subsection{Sensing With Beam Squint Effect Only}\label{sec:sensingSquint}
Let us first consider the wideband beam squint effect only, and the inter-antenna spacing is set as  half the wavelength $d=\frac{\lambda_{c}}{2}$ to avoid beam split effect. 
By designing the value of TTD lines and PSs, we can make the beam squint range adjustable. Specifically, we intend to make the subcarrier beamforming direction varies from the initial angle $\vartheta_{0}$ to the termination angle $\vartheta_{c}$ as  the frequency $f$ increases from $0$ to $F$.
Under wideband assumption, the array gain of the  frequency response vector $\mathbf{g}_{k}$ in \eqref{eq:PSTTDfrequency} and the spatial steering vector $\mathbf{a}(\Theta(f))$ is 
\begin{equation}\label{eq:gainSquint}
\begin{aligned}
&\left|\mathbf{a}^{T}\left(\Theta(f)\right) \mathbf{g}_{k}\right| \\
&=\left|\sum_{m=1}^{M} e^{-j 2 \pi(m-1) (1+\frac{f}{f_c}) \psi _{k}}  e^{j 2 \pi(m-1) \phi_{k}   }  e^{-j2\pi f t_{k,m}} \right|. 
\end{aligned}
\end{equation}
Given the initial angle $\vartheta_{0}$, we have $\Theta(0)=\frac{\sin\vartheta_{0}}{2}$. 
Substituting $\Theta(0)$ back to \eqref{eq:gainSquint}, the array gain at subcarrier frequency $f=0$ is given by 
\begin{equation}
\begin{aligned}
&\left|\mathbf{a}^{T}\left(\Theta(0)\right) \mathbf{g}_{k}\right| =\left|\sum_{m=1}^{M} e^{-j 2 \pi(m-1)  \psi _{k}}  e^{j 2 \pi(m-1) \phi_{k}   }  \right|. 
\end{aligned}
\end{equation}
Hence,  to maximize the array gain at $f=0$, the optimal value of $\phi_{k}$ is $\phi_{k} = \frac{\sin\vartheta_{0}}{2}$. 
Given the termination angle $\vartheta_{c}$ at frequency $f=F$, we have $\Theta(F)=\frac{\sin\vartheta_{c}}{2}(1+\frac{F}{f_c})$. 
Substituting $\Theta(F)$ and $\phi_{k} = \frac{\sin\vartheta_{0}}{2}$ into \eqref{eq:gainSquint}, we have
\begin{equation}
\begin{aligned}
&\left|\mathbf{a}^{T}\left(\Theta(F)\right) \mathbf{g}_{k}\right| \\
&=\left|\sum_{m=1}^{M} e^{-j  \pi(m-1) \left((1+\frac{F}{f_c})\sin\vartheta_{c}-\sin\vartheta_{0}\right)}   e^{-j2\pi F t_{k,m}} \right|. 
\end{aligned}
\end{equation}
To make the array gain achieve the maximum value $M$, one solution of $t_{k,m}$ is 
\begin{equation}\label{eq:sensingTTD}
t_{k, m} =  \frac{1}{2} (m-1) \frac{ \sin \vartheta_{0} - \sin \vartheta_{c} (1 + \frac{F}{f_c}) }{F}.
\end{equation}
Substituting $\phi_{k}$ and $t_{k,m}$ back to \eqref{eq:gainSquint}, we obtain the subcarrier beamforming direction $\vartheta_{k}$ at frequency $f$ as
\begin{equation}\label{eq:sensingangle}
\vartheta_{k} = \arcsin \left( \frac{\sin \vartheta_{0} +  f \frac{  \sin \vartheta_{c} (1 + \frac{F}{f_c}) -\sin \vartheta_{0} }{F} }{1+\frac{f}{f_c} } \right).
\end{equation}
Taking the derivative of $\sin\vartheta_{k}$ with respect to $f$, we have 
\begin{equation}\label{eq:derivative}
\frac{\mathrm{d}\sin\vartheta_{k}}{\mathrm{d}f} =   \frac{\left(\frac{1}{F}+\frac{1}{f_{c}}\right)(\sin \vartheta_{c}-\sin \vartheta_{0})  }{\left(1+\frac{f}{f_c}\right)^{2} } .
\end{equation}
From \eqref{eq:derivative}, we note that the sign of derivative is positive if $\sin \vartheta_{c}>\sin \vartheta_{0}$, and the derivative is negative if $\sin \vartheta_{c}<\sin \vartheta_{0}$. 
Therefore, the beamforming angle $\vartheta_{k}$ of the $k$th RF chain will monotonically  vary from $\vartheta_{0}$ to $\vartheta_{c}$ as the subcarrier frequency $f$ increases.
By steering the beam of each subcarrier towards each direction simultaneously, the BS can sense the users located within the range $[\vartheta_{0},\vartheta_{c}]$.
Then, the user located from the direction $\vartheta_{0}$ to $\vartheta_{c}$ will feedback the subcarrier index with the maximum array gain to the BS, and then the BS utilizes the feedback subcarrier index to calculate the direction of the user according to \eqref{eq:sensingangle}. 

We know that the users are located in the angle range from $-90^{\circ}$ to $90^{\circ}$. 
Hence, the BS can choose $\vartheta_{0}=-90^{\circ}$ and $\vartheta_{c}=90^{\circ}$ to cover the whole range. 
Instead of the whole range, the BS can also sense the user in whichever range from $\vartheta_{0}$ to $\vartheta_{c}$. 
Since the BS has $K$ RF chains, each RF chain can serve one user\footnote{Each user can occupy different bandwidth to avoid the interference.}.
The detailed steps of the user directions sensing are summarized in Algorithm \ref{alg:sensingsquint}.

\begin{algorithm}[!tbp]
\caption{Sensing With Beam Squint Only}
\label{alg:sensingsquint}
\begin{algorithmic}[1]
\STATE{{\bf initialize: } the inter-antenna spacing $d=\frac{\lambda_{c}}{2}$ and the transmission bandwidth $F$.}
\STATE{ Choose the initial angle $\vartheta_{0}$ and calculate $\phi_{k} = \frac{\sin\vartheta_{0}}{2}$. }
\STATE{ Choose the termination angle $\vartheta_{c}$. }
\FOR{ $m=0:M$ }
\STATE{ Set the value of $m$th PS connected to the $k$th RF chain as $e^{j2\pi (m-1) \phi_{k}}$. }
\STATE{ Set the delay of the $m$th TTD line connected to the $k$th RF chain according to \eqref{eq:sensingTTD}. }
\ENDFOR
\STATE{ The $k$th RF chain subcarrier beamforming angle monotonically  varies from $\vartheta_{0}$ to $\vartheta_{c}$. }
\IF{ The user receives the beam } 
\STATE{The subcarrier frequency $f_{d}$ with the maximum array gain is fedback to the BS. }
\STATE{The BS calculates the direction $\vartheta_{k}$ of the user with $f=f_{d}$ according to \eqref{eq:sensingangle}. }
\ELSE
\STATE{The user is not located  in the range from $\vartheta_{0}$ to $\vartheta_{c}$.}
\ENDIF
\ENSURE  The direction $\vartheta_{k}$ of the $k$th user 
\end{algorithmic} 
\end{algorithm}

\subsection{Sensing With Joint Beam Squint and Beam Split}\label{sec:sensingSplit}
As discussed in Section \ref{sec:sensingSquint}, the sensing range is determined by the value of TTD lines. Hence, if the value of TTD lines is  fixed after manufactured, as is usually the case, then the sensing range is not adjustable. To address this issue, the beam split effect can be exploited jointly with the beam squint effect to split the sensing range  into several ranges, which can expand the sensing range and make the sensing range more flexible.

As mentioned in Section \ref{sec:split}, to leverage the beam split effect, the inter-antenna spacing $d$ is larger than $\frac{\lambda_{c}}{2}$. Hence, we assume $d$ is a multiple of $\frac{\lambda_{c}}{2}$ as $d=P\frac{\lambda_{c}}{2}$. 

Similarly, we choose an initial angle $\vartheta_{0}$ and a termination angle $\vartheta_{c}$. 
The array gain is given by 
\begin{equation}\label{eq:gainSquintSplit}
\begin{aligned}
&\left|\mathbf{a}^{T}\left(\Theta(f)\right) \mathbf{g}_{k}\right| \\
&=\left|\sum_{m=1}^{M} e^{-j 2 \pi(m-1) (1+\frac{f}{f_c}) \psi _{k}}  e^{j 2 \pi(m-1) \phi_{k}   }  e^{-j2\pi f t_{k,m}} \right|. 
\end{aligned}
\end{equation}
Given the initial angle $\vartheta_{0}$, we have $\Theta(0)=\frac{d \sin\vartheta_{0}}{\lambda_{c}}$. To maximize the array gain when $f=0$, the optimal value of $\phi_{k}$ is $\phi_{k} = \frac{d \sin\vartheta_{0}}{\lambda_{c}}$. Given the termination angle $\vartheta_{c}$ at frequency $f=F$, we have $\Theta(F)=\frac{d \sin\vartheta_{0}}{\lambda_{c}}(1+\frac{F}{f_c})$. Substituting $\Theta(F)$ and $\phi_{k} = \frac{d \sin\vartheta_{0}}{\lambda_{c}}$ into \eqref{eq:gainSquint}, we have
\begin{equation}\label{eq:TTDsquintsplit}
t_{k, m} =  \frac{d}{\lambda_{c}} (m-1) \frac{ \sin \vartheta_{0} - \sin \vartheta_{c} (1 + \frac{F}{f_c}) }{F}.
\end{equation}
Given $\phi_{k}$ and $t_{k,m}$, the $k$th RF chain subcarrier beamforming angle monotonically varies from $\vartheta_{0}$ to $\vartheta_{c}$ as the subcarrier frequency increases. 
The user located between $\vartheta_{0}$ to $\vartheta_{c}$ will receive the beam and feedback the  subcarrier index with the maximum array gain to the BS.
Then the BS can obtain the user angle to maximize \eqref{eq:gainSquintSplit} with the feedback frequency $f$, which yields 
\begin{equation}\label{eq:sensingangleSquintSplit}
\sin\vartheta_{k} =  \frac{\sin \vartheta_{0} -  f \frac{  \lambda_{c} }{d} \frac{t_{k, m}}{m-1} }{1+\frac{f}{f_c} } .
\end{equation}

\begin{figure}[!tbp]
\centering
\includegraphics[width=0.4\textwidth]{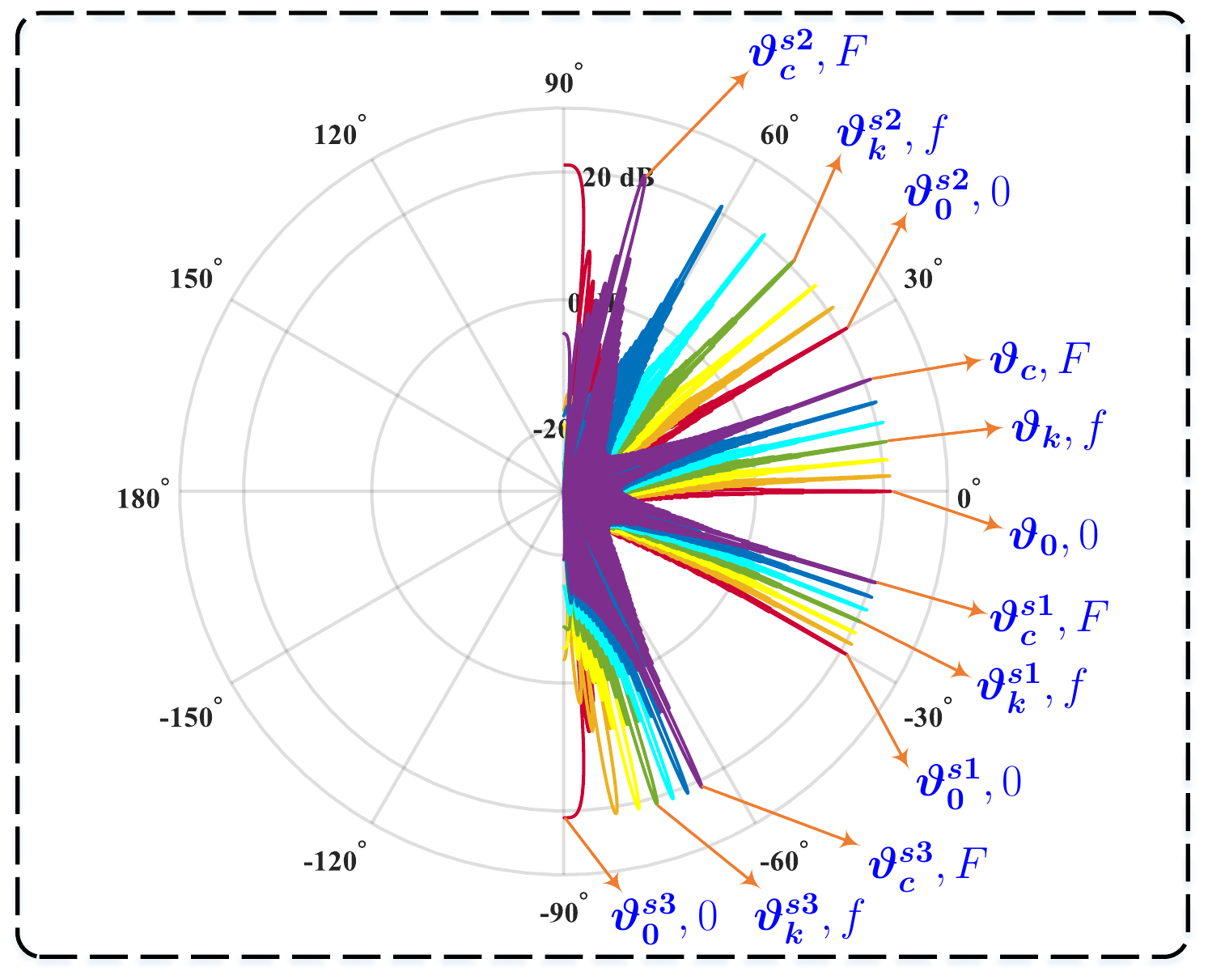}
\caption{The beam pattern of the sensing method with joint beam squint and beam split, where $M=128$, $d=2\frac{\lambda_{c}}{2}$ and the same color denotes the splitted beams at one frequency point.}
\label{fig:SplitTTD}
\end{figure}
However, as a consequence of the beam split effect, the angle $\vartheta_{0}$ to $\vartheta_{c}$ will split into several distinct angles, which is shown in Fig. \ref{fig:SplitTTD}. 
The splitted angles corresponding to $\vartheta_{0}$ are denoted by $[\vartheta_{0}^{s1},\vartheta_{0}^{s2},\vartheta_{0}^{s3},\cdots]$, and the splitted angles for $\vartheta_{c}$ are denoted by $[\vartheta_{c}^{s1},\vartheta_{c}^{s2},\cdots]$.
Similarly, the angle in \eqref{eq:sensingangleSquintSplit} at frequency $f$ will split into $[\vartheta_{k}^{s1},\vartheta_{k}^{s2},\cdots]$.
Note that the number of the splitted angles is not necessarily one (determined by the number $P$ and \eqref{eq:splitwideband}).
Nevertheless, for ease of notation, we assume there is only one splitted angle and abuse $\vartheta_{0}^{s}$, $\vartheta_{c}^{s}$ and $\vartheta_{k}^{s}$ to denote the splitted angles of $\vartheta_{0}$, $\vartheta_{c}$ and $\vartheta_{k}$, respectively. 
The value of $\vartheta_{0}^{s}$ and $\vartheta_{c}^{s}$ can be obtained according to \eqref{eq:splitwideband}.
Given $\vartheta_{0}^{s}$ and $t_{k,m}$, we can obtain 
\begin{equation}\label{eq:anglesensingsplit}
\sin\vartheta_{k}^{s} =  \frac{\sin \vartheta_{0}^{s} -  f \frac{  \lambda_{c} }{d} \frac{t_{k, m}}{m-1} }{1+\frac{f}{f_c} } .
\end{equation}
With the feedback frequency $f$, the BS will obtain $\vartheta_{k}$ and the splitted $\vartheta_{k}^{s}$. 
However, the BS can not determine which one in $\{\vartheta_{k}, \vartheta_{k}^{s}\}$ is the true angle of the user.
Therefore, we should eliminate the ambiguity caused by the beam split effect to  find the true direction from $\{\vartheta_{k}, \vartheta_{k}^{s}\}$. 

Next, we propose a solution  to eliminate the ambiguity. 
Specifically, since we know that the true angle is in the set $\{\vartheta_{k}, \vartheta_{k}^{s}\}$, we can construct another validation set $\{\tilde\vartheta_{k}, \tilde\vartheta_{k}^{s}\}$ whose elements are different from the set $\{\vartheta_{k}, \vartheta_{k}^{s}\}$ except for the true angle. 
Then, taking the intersection of the two sets  $\{\vartheta_{k}, \vartheta_{k}^{s}\}\cap \{\tilde\vartheta_{k}, \tilde\vartheta_{k}^{s}\}$, we can obtain the true angle of the user. 
We name the above method that eliminates the ambiguity as \emph{intersection validation}.
To construct the validation set $\{\tilde\vartheta_{k}, \tilde\vartheta_{k}^{s}\}$, we only need to change the initial angle $\vartheta_{0}$ into $\tilde\vartheta_{0}$ but keep the delay of the TTD lines unchanged. 
Consequently, the termination angle $\vartheta_{c}$ will change into $\tilde\vartheta_{c}$. 
Additionally, we should guarantee that this new range from $\tilde\vartheta_{0}$ to $\tilde\vartheta_{c}$ still cover the user to be sensed.
Then, the subcarrier beamforming angle monotonically varies in the new range from $\tilde\vartheta_{0}$ to $\tilde\vartheta_{c}$. 
The user will receive the beam and feedback a new subcarrier frequency $\omega$ with  the maximum array gain. 
Based on the feedback frequency $\omega$, the BS can calculate the angles in the validation set $\{\tilde\vartheta_{k}, \tilde\vartheta_{k}^{s}\}$.
Next, we will find the condition that the validation set $\{\tilde\vartheta_{k}, \tilde\vartheta_{k}^{s}\}$ are different from the set $\{\vartheta_{k}, \vartheta_{k}^{s}\}$ except the true angle.

Specifically, we change the initial angle into $\tilde\vartheta_{0}$, and we have $\Theta(0)=\frac{d \sin\tilde\vartheta_{0}}{\lambda_{c}}$. 
Then, the optimal value of $\phi_{k}$ is $\phi_{k} = \frac{d \sin\tilde\vartheta_{0}}{\lambda_{c}}$. 
While, the delay $t_{k, m}$ of the TTD lines is unchanged as given by \eqref{eq:TTDsquintsplit}.
With the changed initial angle $\tilde\vartheta_{0}$ and unchanged $t_{k, m}$, the termination angle will change into $\tilde\vartheta_{c}$, which yields
\begin{equation}\label{eq:changedtermination}
\sin\tilde\vartheta_{c} =  \frac{\sin \tilde\vartheta_{0} -  F \frac{  \lambda_{c} }{d} \frac{t_{k, m}}{m-1} }{1+\frac{F}{f_c} } .
\end{equation}
Given the new initial angle $\tilde\vartheta_{0}$ and the termination angle $\tilde\vartheta_{c}$, the user can receive the subcarrier beamforming signal and feedback the different subcarrier frequency $\omega$ with  the maximum array gain. 
Then, the BS calculate the direction of the user as 
\begin{equation}\label{eq:changedangle}
\sin\tilde\vartheta_{k} =  \frac{\sin \tilde\vartheta_{0} -  \omega \frac{  \lambda_{c} }{d} \frac{t_{k, m}}{m-1} }{1+\frac{\omega}{f_c} } .
\end{equation}
According to \eqref{eq:splitwideband}, we can obtain the splitted angle of $\tilde\vartheta_{0}$ and $\tilde\vartheta_{c}$ as $\tilde\vartheta_{0}^{s}$ and $\tilde\vartheta_{c}^{s}$, respectively. Given $\tilde\vartheta_{0}^{s}$ and $t_{k, m}$, we can calculate the splitted angle of $\tilde\vartheta_{k}$ as 
\begin{equation}\label{eq:changedsplitangle}
\sin\tilde\vartheta_{k}^{s} =  \frac{\sin \tilde\vartheta_{0}^{s} -  \omega \frac{  \lambda_{c} }{d} \frac{t_{k, m}}{m-1} }{1+\frac{\omega}{f_c} } .
\end{equation}

We assume that the true angle of the user in the set $\{\vartheta_{k}, \vartheta_{k}^{s}\}$ and the validation set $\{\tilde\vartheta_{k}, \tilde\vartheta_{k}^{s}\}$ is identical, which yields
\begin{equation}\label{eq:equationSquintSplit}
\begin{aligned}
 \sin\vartheta_{k} &= \sin\tilde\vartheta_{k} \\
 \frac{\sin \vartheta_{0} -  f \frac{  \lambda_{c} }{d} \frac{t_{k, m}}{m-1} }{1+\frac{f}{f_c} } &= \frac{\sin \tilde\vartheta_{0} -  \omega \frac{  \lambda_{c} }{d} \frac{t_{k, m}}{m-1} }{1+\frac{\omega}{f_c} }.
\end{aligned}
\end{equation}
From the fact that $\frac{x+a}{y+a}=\frac{c+bx}{d+by} \Rightarrow \frac{x+a}{y+a}=\frac{c-ba}{d-ba}$, we can simplify \eqref{eq:equationSquintSplit} as
\begin{equation}\label{eq:ambiguity1}
 \frac{ f_{c} + \omega }{ f_{c} + f } = \frac{ \sin \tilde\vartheta_{0} +  f_{c} \frac{  \lambda_{c} }{d} \frac{t_{k, m}}{m-1} }{\sin \vartheta_{0} +  f_{c} \frac{  \lambda_{c} }{d} \frac{t_{k, m}}{m-1} }.
\end{equation}
Additionally, we suppose that the splitted angle in the set $\{\vartheta_{k}, \vartheta_{k}^{s}\}$ and the validation set $\{\tilde\vartheta_{k}, \tilde\vartheta_{k}^{s}\}$ is also identical, which indicates we can not eliminate the ambiguity by intersection validation. Then, we have
\begin{equation}\label{eq:equation2SquintSplit}
\begin{aligned}
 \sin\vartheta_{k}^{s} &= \sin\tilde\vartheta_{k}^{s} \\
 \frac{\sin \vartheta_{0}^{s} -  f \frac{  \lambda_{c} }{d} \frac{t_{k, m}}{m-1} }{1+\frac{f}{f_c} } &= \frac{\sin \tilde\vartheta_{0}^{s} -  \omega \frac{  \lambda_{c} }{d} \frac{t_{k, m}}{m-1} }{1+\frac{\omega}{f_c} }.
\end{aligned}
\end{equation}
Given \eqref{eq:equationSquintSplit} and \eqref{eq:equation2SquintSplit}, and after some algebra, we obtain 
\begin{equation}\label{eq:ambiguity2}
\sin \tilde\vartheta_{0} - \sin \tilde\vartheta_{0}^{s} = \frac{ f_{c} + \omega }{ f_{c} + f } \left( \sin \vartheta_{0} - \sin \vartheta_{0}^{s} \right).
\end{equation}
From \eqref{eq:ambiguity1} and \eqref{eq:ambiguity2}, we have
\begin{equation}\label{eq:ambiguity}
 \frac{ \sin \tilde\vartheta_{0} +  f_{c} \frac{  \lambda_{c} }{d} \frac{t_{k, m}}{m-1} }{\sin \vartheta_{0} +  f_{c} \frac{  \lambda_{c} }{d} \frac{t_{k, m}}{m-1} } = \frac{\sin \tilde\vartheta_{0} - \sin \tilde\vartheta_{0}^{s}}{\sin \vartheta_{0} - \sin \vartheta_{0}^{s}}.
\end{equation}
If \eqref{eq:ambiguity} is satisfied, then the ambiguity can not be eliminated by the intersection validation since the  splitted angle in the set in the set $\{\vartheta_{k}, \vartheta_{k}^{s}\}$ and the validation set $\{\tilde\vartheta_{k}, \tilde\vartheta_{k}^{s}\}$ is also identical. 
In other word, the BS should choose the value of $\vartheta_{0}$ and  $\tilde\vartheta_{0}$ to  make the equation \eqref{eq:ambiguity} not satisfied, and hence only the true angle of the user  in the set $\{\vartheta_{k}, \vartheta_{k}^{s}\}$ and the validation set $\{\tilde\vartheta_{k}, \tilde\vartheta_{k}^{s}\}$ is identical. 
Then,  the true direction of the $k$th user can be obtained by  the intersection of the two sets  $\{\vartheta_{k}, \vartheta_{k}^{s}\}\cap \{\tilde\vartheta_{k}, \tilde\vartheta_{k}^{s}\}$.

It is worth noting that as a consequence of beam split effect, the sensing range from $\vartheta_{0}$ to $\vartheta_{c}$ has several splitted ranges from $\vartheta_{0}^{s}$ to $\vartheta_{c}^{s}$. If there are overlaps between the sensing range and the splitted ranges, then more uncertainty and ambiguity will occur during sensing.  
Therefore, the value of $\vartheta_{0}$ and $\vartheta_{c}$ should be elaborately chosen to avoid overlaps among these splitted ranges. 
For example, in Fig. \ref{fig:SplitTTD}, no overlaps occur among these splitted ranges as expected.  

Similarly, since the BS has $K$ RF chains, each RF chain can serve one user.
The detailed steps of the user directions sensing with joint beam squint and beam split are summarized in Algorithm \ref{alg:sensingsquintsplit}. 

\begin{algorithm}[!tbp]
\caption{Sensing With Joint Beam Squint And Beam Split}
\label{alg:sensingsquintsplit}
\begin{algorithmic}[1]
\STATE{{\bf initialize: } the inter-antenna spacing $d=P\frac{\lambda_{c}}{2}$ and the transmission bandwidth $F$.}
\WHILE{  Overlaps occur   }
\STATE Choose an initial angle $\vartheta_{0}$.
\STATE{ Choose a termination angle $\vartheta_{c}$. }
\STATE Obtain the splitted angles $\vartheta_{0}^{s}$ and $\vartheta_{c}^{s}$ via \eqref{eq:splitwideband}.
\IF{ No overlaps between $[\vartheta_{0},\vartheta_{c}]$ and $[\vartheta_{0}^{s},\vartheta_{c}^{s}]$ }
\STATE{\bf break}
\ENDIF
\ENDWHILE
\FOR{ $m=0:M$ }
\STATE{ Set the value of $m$th PS connected to the $k$th RF chain as $e^{j2\pi (m-1) \phi_{k}}$, where $\phi_{k} = \frac{d \sin\vartheta_{0}}{\lambda_{c}}$. }
\STATE{ Set the delay of the $m$th TTD line connected to the $k$th RF chain according to \eqref{eq:TTDsquintsplit}. }
\ENDFOR
\STATE{ The $k$th RF chain subcarrier beamforming angle monotonically  varies from $\vartheta_{0}$ to $\vartheta_{c}$, and the splitted beamforming angle varies from  $\vartheta_{0}^{s}$ to $\vartheta_{c}^{s}$. }
\IF{ The user receives the beam } 
\STATE{The subcarrier frequency $f_{d}$ with the maximum array gain is fedback to the BS. }
\STATE{ According to \eqref{eq:sensingangleSquintSplit} and \eqref{eq:anglesensingsplit}, the BS calculates $\vartheta_{k}$ and $\vartheta_{k}^{s}$ with $f=f_{d}$. }
\ELSE
\STATE{The user is located neither in the range from $\vartheta_{0}$ to $\vartheta_{c}$ nor in the range from $\vartheta_{0}^{s}$ to $\vartheta_{c}^{s}$.  }
\STATE Terminate this algorithm.
\ENDIF
\WHILE{  Overlaps occur or the user receives no beams }
\STATE Change the initial angle as $\tilde\vartheta_{0}$.
\STATE{ Calculate the changed termination angle $\tilde\vartheta_{c}$ via \eqref{eq:changedtermination}. }
\STATE Obtain the splitted angles $\tilde\vartheta_{0}^{s}$ and $\tilde\vartheta_{c}^{s}$ via \eqref{eq:splitwideband}.
\IF{ $[\tilde\vartheta_{0},\tilde\vartheta_{c}]$ and $[\tilde\vartheta_{0}^{s},\tilde\vartheta_{c}^{s}]$ have no overlaps and the user receives beams }
\STATE{\bf break}
\ENDIF
\ENDWHILE
\FOR{ $m=0:M$ }
\STATE{ Change the value of $m$th PS connected to the $k$th RF chain as $e^{j2\pi (m-1) \phi_{k}}$, where $\phi_{k} = \frac{d \sin\tilde\vartheta_{0}}{\lambda_{c}}$. }
\ENDFOR
\STATE{ The $k$th RF chain subcarrier beamforming angle monotonically  varies from $\tilde\vartheta_{0}$ to $\tilde\vartheta_{c}$, and the splitted beamforming angle varies from  $\tilde\vartheta_{0}^{s}$ to $\tilde\vartheta_{c}^{s}$. } 
\STATE{The user feedbacks subcarrier frequency $\omega_{d}$ with the maximum array gain to the BS. }
\STATE{ According to \eqref{eq:changedangle} and \eqref{eq:changedsplitangle}, the BS calculates $\tilde\vartheta_{k}$ and $\tilde\vartheta_{k}^{s}$ with $\omega=\omega_{d}$. }
\STATE The user's angle $\vartheta_{k}$  is obtained via $\{\vartheta_{k}, \vartheta_{k}^{s}\}\cap \{\tilde\vartheta_{k}, \tilde\vartheta_{k}^{s}\}$.
\ENSURE  The direction $\vartheta_{k}$ of the $k$th user 
\end{algorithmic} 
\end{algorithm}

\subsection{The Number of MIMO-OFDM Subcarriers}
It is worth pointing out that in Section \ref{sec:sensingSquint} and \ref{sec:sensingSplit}, we assume that the BS implements sensing algorithms with off-grid subcarrier frequency, i.e.,  
the user can receive the beam at any frequency in the transmission bandwidth $F$. 
However, in OFDM system, the subcarrier frequency is on-grid with a uniform spacing.
The whole sensing range is then gridded into $N$ points with total $N$ OFDM subcarriers. 
As illustrated in Fig. \ref{fig:SplitTTD}, the sensing range is grided into $7$ points.
Hence, the BS can only sense the  on-grid directions corresponding to the OFDM uniformly spaced subcarrier, and consequently the subcarrier number $N$ will influence the sensing accuracy.
Additionally, we note that the subcarrier beamforming angle is an arcsine function of the subcarrier frequency $f$, as shown  in \eqref{eq:sensingangle} and \eqref{eq:sensingangleSquintSplit}.
Consequently, even though the subcarrier frequency is equispaced, the corresponding subcarrier beamforming angle is not uniformly spaced.
Nevertheless, traditional beam sweeping methods also only operate in this on-grid manner \cite{WJtraining,DJtraning,WJHtraining,NBYtraining}.
Hence, to improve the sensing accuracy, increasing the grid points is rational and indispensable.

The traditional beam sweeping methods perform over different time slots, i.e., the whole range is discretized into $Q$ directions, and the BS emits beams by $Q$ times to steer towards the directions individually.
Hence, it is time-consuming to achieve higher accuracy with larger $Q$.
While, the proposed sensing methods operate over frequency-domain, and all the gridded directions  are simultaneously covered by the beams of the subcarriers within only one OFDM block.
Therefore, the proposed sensing method can reduce the time overhead significantly.

For the sensing method with beam squint in Section \ref{sec:sensingSquint}, the number of OFDM subcarriers $N$ should be sufficiently large if the sensing range from $\vartheta_{0}$ to $\vartheta_{c}$ is wide. However, for the systems with limited number of subcarriers, one compromise  way to provide a high sensing accuracy is to divide the  whole range from $\vartheta_{0}=-90^{\circ}$ to $\vartheta_{c}=90^{\circ}$ into several  ranges. Then, the BS can deal with the divided ranges individually. 
This compromise requires more time overhead indeed since the BS need to sense the users several times. 
Nevertheless, by adjusting the sensing range according to the density of the subcarriers beams, this compromise can also address the issue of not uniformly spaced subcarrier beamforming angle. 

For the sensing technique with joint beam squint and beam split in Section \ref{sec:sensingSplit}, influences of the wide sensing range and the non-uniformly spaced subcarrier beam are unremarkable.
Since the sensing range can  split to several ranges, we can choose the inter-antenna spacing to make the number of the splitted ranges as large as possible.
Additionally, the overlaps among the splitted ranges are prevented elaborately. 
Then, each splitted range is sufficiently narrow, and the sensing accuracy can be improved with limited number of subcarriers.
According to the beam split effect, we note that the splitted beams are transmitted simultaneously. 
Therefore, the users in the splitted ranges can be sensed simultaneously. 
As mentioned in Section \ref{sec:sensingSplit}, the main time overhead results from the intersection validation since the BS need to change the sensing initial angle and sense the user one more time.
Nevertheless, this method only need to sense the users twice with intersection validation. 
Compared with the beam squint-based sensing method with several divided ranges, this joint beam squint and beam split-based  technique can also provide high sensing accuracy with limited $N$ subcarriers.

\section{Numerical Results} \label{sec:sim}
In this section, we present simulation results to show the effectiveness of the proposed sensing methods, and compare the two proposed methods. We consider a mmWave wideband transmission scenario with $f_{c}=30$ GHz and $F=6$ GHz.

Under the assumption of wideband transmission, the SNR (in dB) is defined as 
\begin{equation}
\mathrm{SNR} \triangleq 10\log_{10}\left( \frac{\| \mathbf{a}(\Theta(f)) \|^{2}_{2}}{\sigma^2} \right),
\end{equation}
where the Gaussian noise variance is set as $\sigma^2=1$. The performance metric of  the sensing method is the root mean
square error (RMSE) defined as
\begin{equation} 
\mathrm{RMSE} \triangleq \sqrt{ \mathbb{E}\left\{ \frac{\sum_{k=1}^{K}\left( \hat\vartheta_{k}  - \vartheta_{k} \right)^2}{K} \right\} },
\end{equation}
where $\hat\vartheta_{k}$ is the sensed user direction and $\vartheta_{k}$ is the ground truth angle.

The performances of the following algorithms will be taken as comparisons:
\begin{itemize}
\item ``squint, AoD $\in[-80^{\circ},80^{\circ}]$, range $\in[-80^{\circ},80^{\circ}]$'': the sensing method with beam squint only, where the directions are generated within $[-80^{\circ},80^{\circ}]$, and the sensing range is from $-80^{\circ}$ to $80^{\circ}$.
\item ``squint, AoD $\in[60^{\circ},80^{\circ}]$, range $\in[-80^{\circ},80^{\circ}]$'': the sensing method with beam squint only, where the directions are generated within $[60^{\circ},80^{\circ}]$, and the sensing range is from $-80^{\circ}$ to $80^{\circ}$.
\item ``squint, AoD $\in[0^{\circ},20^{\circ}]$, range $\in[-80^{\circ},80^{\circ}]$'': the sensing method with beam squint only, where the angles are generated within $[0^{\circ},20^{\circ}]$, and the sensing range is from $-80^{\circ}$ to $80^{\circ}$.
\item ``squint, AoD \& range $\in[0^{\circ},20^{\circ}]$'': the sensing method with beam squint only, where both 
the generated angles and the sensing range are within $[0^{\circ},20^{\circ}]$.
\item ``squint, AoD \& range $\in[60^{\circ},80^{\circ}]$'': the sensing method with beam squint only, where both 
the generated angles and the sensing range are within $[60^{\circ},80^{\circ}]$.
\item ``squint \& split, $\frac{d}{\lambda_{c}}=\#$'': the sensing method with joint beam squint and beam split, where the inter-antenna spacing is $d=\# \cdot \lambda_{c}$, and the directions are generated within $[-80^{\circ},80^{\circ}]$.
\end{itemize}

\begin{figure}[!tbp]
\centering
\includegraphics[width=0.4\textwidth]{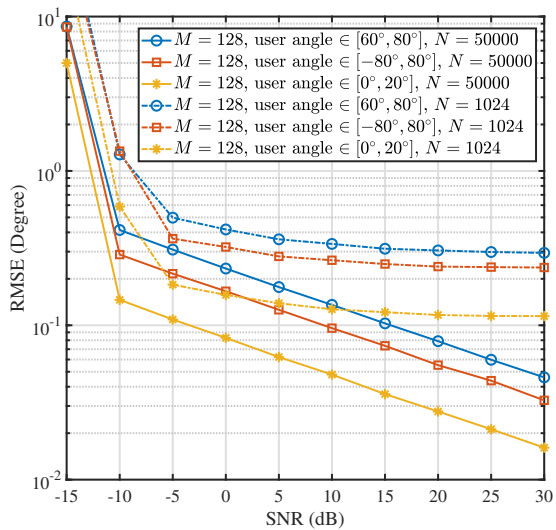}
\caption{$\mathrm{RMSE}$ of the sensing methods with beam squint only versus SNR, where $M=128$.}
\label{fig:RMSEvsSNRsquint}
\end{figure}
In Fig. \ref{fig:RMSEvsSNRsquint}, the RMSE of the sensing method with beam squint only is plotted over SNR, where $M=128$. 
Note that  the sensing range of all the algorithms in Fig. \ref{fig:RMSEvsSNRsquint} is from $-80^{\circ}$ to $80^{\circ}$, while the subcarriers number $N$ and the angles distribution are different.   
It can be found that the sensing accuracy will be improved by increasing the number of subcarriers $N$.
The reason is that the sensing method is in an on-grid manner corresponding to the OFDM uniformly spaced subcarrier.
With total $N$ OFDM subcarriers, the sensing range is gridded into $N$ points.
Therefore, smaller subcarriers number $N$ will result in a thinner grid and limit the sensing accuracy.
In Fig. \ref{fig:RMSEvsSNRsquint}, it is seen that there are nonzero error floors for the algorithms with $N=1024$, whereas the error floors not occur for $N=50000$, which means the sensing accuracy will be limited by a thinner grid.
Additionally, with the same number $N$, the sensing performance varies for different  directions distributions. 
Seen from Fig. \ref{fig:RMSEvsSNRsquint}, with the same number $N$, the performance for the angles generated within $[0^{\circ},20^{\circ}]$ is better than the case for the angles generated within $[60^{\circ},80^{\circ}]$. 
The reason is that according to \eqref{eq:sensingangle}, the subcarrier beamforming angle is a non-linear arcsine function of the subcarrier frequency $f$.
Even though the OFDM subcarrier is equispaced, the corresponding subcarrier beamforming angle is not uniformly spaced. 
Hence, the sensing range is not uniformly gridded, and the grid is more linear and uniform within $[0^{\circ},20^{\circ}]$, which results in higher sensing accuracy.  

\begin{figure}[!tbp]
\centering
\includegraphics[width=0.4\textwidth]{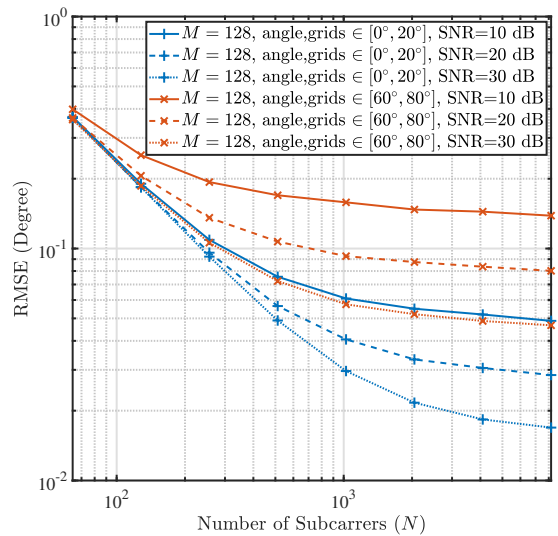}
\caption{$\mathrm{RMSE}$ of the sensing method with beam squint only versus the subcarriers number $N$, where $M=128$.}
\label{fig:RMSEvsNsquint}
\end{figure}
Fig. \ref{fig:RMSEvsNsquint} shows the RMSE of the sensing method with beam squint only versus $N$, where  $M=128$. The sensing range and the AoDs  distribution are the same within $[0^{\circ},20^{\circ}]$ as well as $[60^{\circ},80^{\circ}]$.
We observe that  RMSE decreases significantly in the small $N$ regime, whereas  RMSE tends to converge to a nonzero error floor in the large $N$ regime. 
Besides, the nonzero error floor decreases as SNR increases.
This result shows that increasing the subcarriers number $N$ can improve sensing accuracy significantly in the small $N$ regime, while  SNR becomes the performance bottleneck if $N$ is sufficiently large.
Moreover, we see that the performance for the range $[0^{\circ},20^{\circ}]$ is better the range $[60^{\circ},80^{\circ}]$ for a given number $N$ and the same SNR, which indicates that the sensing range is not uniformly gridded and the grid is more even within $[0^{\circ},20^{\circ}]$.

\begin{figure}[!tbp]
\centering
\includegraphics[width=0.4\textwidth]{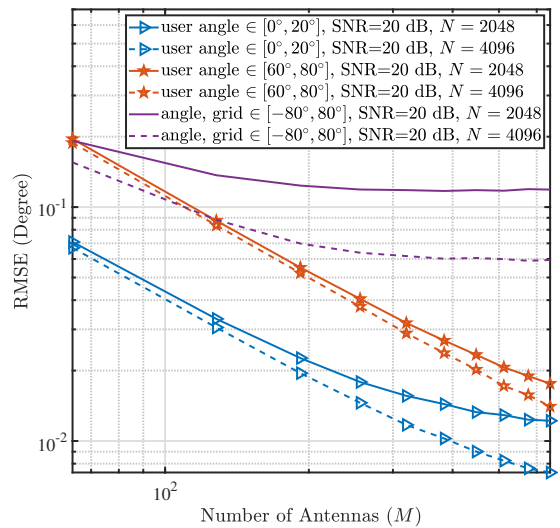}
\caption{$\mathrm{RMSE}$ of the sensing method with beam squint only versus $M$, where SNR is $20$ dB.}
\label{fig:RMSEvsMsquint}
\end{figure}
Fig. \ref{fig:RMSEvsMsquint} displays the RMSE of the sensing method with beam squint only versus $M$, where SNR is $20$ dB.
It is seen in Fig. \ref{fig:RMSEvsMsquint} that  RMSE decreases as $M$ increases for all algorithms. 
According to the array signal processing theory, the beam width (the width of the main lobe) is the reciprocal of $2M$.
Hence, by increasing the antenna number $M$, the BS can focus energy towards a narrow width to improve sensing accuracy.
Nevertheless, when $M$ is sufficiently large,  RMSE will converge to a nonzero error floor, and this error floor will decrease as the subcarriers number $N$ increase.
This result indicates that the subcarriers number becomes one of  the performance bottlenecks if $M$ is sufficiently large. 
In addtion, as we can see in Fig. \ref{fig:RMSEvsMsquint},  the performance for the range $[0^{\circ},20^{\circ}]$ is better than the range $[60^{\circ},80^{\circ}]$ and $[-80^{\circ},80^{\circ}]$, which indicates that the sensing range is more linearly and uniformly gridded in $[0^{\circ},20^{\circ}]$.
Furthermore, we observe that the performance for the range $[-80^{\circ},80^{\circ}]$ is notably poor than the range $[0^{\circ},20^{\circ}]$ and $[60^{\circ},80^{\circ}]$. 
The reason is straightforward that the the grid in the sensing range $[-80^{\circ},80^{\circ}]$ is much thinner with the same $N$. 
Thus, one way to improve the performance for the range $[-80^{\circ},80^{\circ}]$ is providing a denser grid by increasing the subcarriers number  $N$. 
Otherwise, with limited $N$, a compromise way is dividing $[-80^{\circ},80^{\circ}]$ into several narrow ranges and sensing in the divided ranges individually.
This compromise yields more time overhead.

\begin{figure}[!tbp]
\centering
\includegraphics[width=0.4\textwidth]{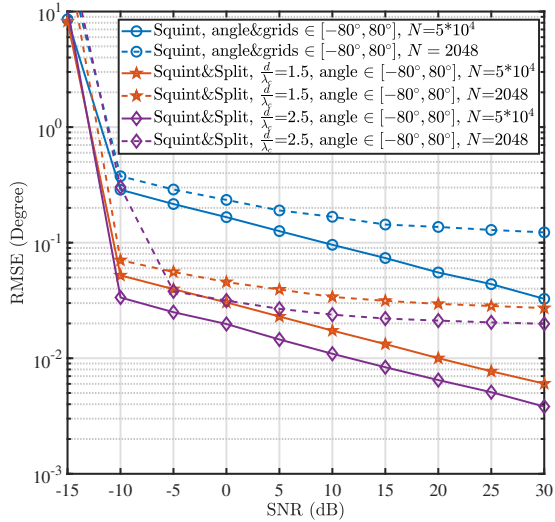}
\caption{$\mathrm{RMSE}$ of the sensing method with joint beam squint and beam split versus SNR, where $M=128$.}
\label{fig:RMSEvsSNRsplit}
\end{figure}
Next, we study the performance of the sensing method with joint beam squint and beam split.  Fig. \ref{fig:RMSEvsSNRsplit} plots the RMSE of the sensing method with joint beam squint and beam split versus SNR, where $M=128$.
Similar to the results in Fig. \ref{fig:RMSEvsSNRsquint}, the RMSE of the sensing method with joint beam squint and beam split decreases as SNR increases, but a nonzero error floor occurs with smaller number $N$.
This result means that in high SNR regime, the subcarriers number $N$ is the performance bottleneck.
In Fig. \ref{fig:RMSEvsSNRsplit}, the directions are generated in the range $[-80^{\circ},80^{\circ}]$ for both the sensing method with beam squint only as well as the sensing method with joint beam squint and beam split.
Compared with the sensing method with beam squint only, the sensing method with joint beam squint and beam split has remarkable performance improvement with the same number $N$.
Since the sensing range is splitted into several ranges with beam split effect, each range is much narrow than $[-80^{\circ},80^{\circ}]$, and the grid in each splitted range is much dense.
Hence,  the sensing method with joint beam squint and beam split has superior performance.
Besides, we see that the performance has been improved with larger $\frac{d}{\lambda_{c}}$.
By increasing the inter-antenna spacing, the sensing range will be splitted into more ranges.
Then, each splitted sensing range will be narrower, and the grid becomes denser, which improve the sensing performance.

\begin{figure}[!tbp]
\centering
\includegraphics[width=0.4\textwidth]{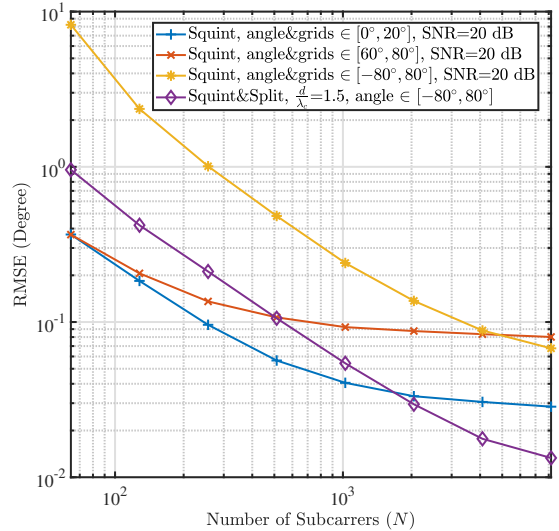}
\caption{$\mathrm{RMSE}$ of the sensing method with joint beam squint and beam split versus $N$, where $M=128$.}
\label{fig:RMSEvsNsplit}
\end{figure}
In Fig. \ref{fig:RMSEvsNsplit}, the RMSE is plotted versus $N$, where $M=128$ and SNR is $20$ dB. 
Similarly, RMSE of the method with joint beam squint and beam split decreases as $N$ increases, since larger $N$ provides denser  grid in each splitted sensing range.
Additionally, comparisons between the sensing method with  joint beam squint and beam split as well as the method with beam squint only are shown in Fig. \ref{fig:RMSEvsNsplit}.
It is seen that the method with joint beam squint and beam split has superior performance over the method with beam squint only for the sensing range $[-80^{\circ},80^{\circ}]$.
While for the sensing range $[0^{\circ},20^{\circ}]$ and  $[60^{\circ},80^{\circ}]$, these two methods have close performance. 
It means that only with the compromise to divide the sensing range into several narrow ranges, the sensing method with beam squint only can outperforms the method with  joint beam squint and beam split.
However, this compromise will lead to higher time overhead for the method with beam squint only, since the divided narrow ranges need to be sensed one by one.
This result shows the superiority of the method with  joint beam squint and beam split.
Moreover, similar conclusion can be found that RMSE tends to converge to a nonzero error floor as $N$ increases, which indicates that SNR and other factors become the performance bottleneck.

\begin{figure}[!tbp]
\centering
\includegraphics[width=0.4\textwidth]{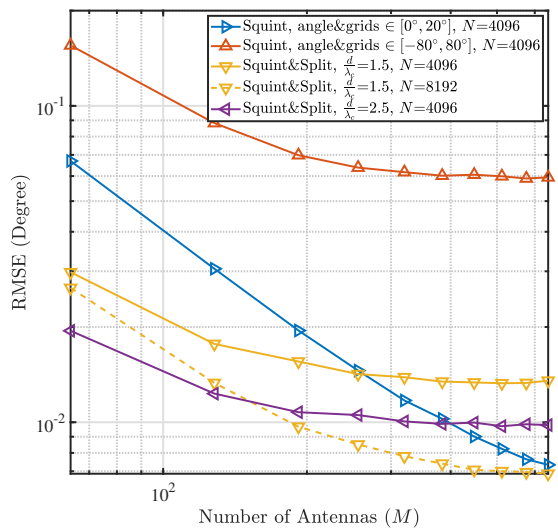}
\caption{$\mathrm{RMSE}$ of the sensing method with joint beam squint and beam split versus $M$, where SNR is $20$ dB.}
\label{fig:RMSEvsMsplit}
\end{figure}
Fig. \ref{fig:RMSEvsMsplit} displays the RMSE  of the sensing method with joint beam squint and beam split versus $M$, where SNR is $20$ dB. 
Similarly, we observe that increasing $M$ can improve the sensing accuracy in the small $M$ regime, and the error floor occurs if $M$ is sufficiently large.
Additionally, as can be seen, when the sensing range is divided into narrow ones (e.g., $[0^{\circ},20^{\circ}]$), the sensing method with beam squint only has superior performance over the method with joint beam squint and beam split.
This yields the superiority of the method with  joint beam squint and beam split. 	
Furthermore, we see that the performance of the method with  joint beam squint and beam split can be further improved by increasing the inter-antenna spacing $d$.
Since increasing $d$ produces more splitted sensing ranges, the grid in each splitted sensing range becomes denser, which results in performance improvement.

\section{Conclusions}\label{sec:con}
In this paper, we propose a novel ISAC technique with joint beam squint and beam split for massive MIMO-OFDM systems.
Specifically, instead of mitigating the beam squint effect from large antenna array and wide transmission bandwidth, we take advantage of the beam squint to make different subcarriers emit beams towards different directions simultaneously.
The users then feedback the subcarrier frequency  with the maximum array gain to the BS.
Then, the BS obtain the direction of the user based on the feedback frequency.
Moreover, the beam split effect which is caused by increasing the inter-antenna spacing is exploited to expand the sensing range.
Numerical results have been carried out to show the effectiveness and the high performance of the proposed method in user directions sensing.


\bibliographystyle{IEEEtran}
\bibliography{IEEEabrv,ISACSquintSplit}

\end{document}